\input harvmac
\input amssym

\baselineskip 13pt

\lref\LarsenXM{
  F.~Larsen,
  ``The Perturbation spectrum of black holes in N=8 supergravity,''
Nucl.\ Phys.\ B {\bf 536}, 258 (1998).
[hep-th/9805208].
}
\lref\MichelsonKN{
  J.~Michelson and M.~Spradlin,
  ``Supergravity spectrum on AdS(2) x S**2,''
JHEP {\bf 9909}, 029 (1999).
[hep-th/9906056].
}
\lref\LeeYU{
  J.~Lee and S.~Lee,
  ``Mass spectrum of D = 11 supergravity on AdS(2) x S**2 x T**7,''
Nucl.\ Phys.\ B {\bf 563}, 125 (1999).
[hep-th/9906105].
}
\lref\CorleyUZ{
  S.~Corley,
  ``Mass spectrum of N=8 supergravity on AdS(2) x S**2,''
JHEP {\bf 9909}, 001 (1999).
[hep-th/9906102].
}

\lref\SalamXD{
  A.~Salam and J.~A.~Strathdee,
  ``On Kaluza-Klein Theory,''
Annals Phys.\  {\bf 141}, 316 (1982).
}

\lref\GrumillerNM{
  D.~Grumiller, W.~Kummer and D.~V.~Vassilevich,
Phys.\ Rept.\  {\bf 369}, 327 (2002).
[hep-th/0204253].
}

\lref\AbrikosovNJ{
  A.~A.~Abrikosov, Jr.,
  ``Fermion states on the sphere S**2,''
Int.\ J.\ Mod.\ Phys.\ A {\bf 17}, 885 (2002).
[hep-th/0111084].
}

\lref\VassilevichXT{
  D.~V.~Vassilevich,
  ``Heat kernel expansion: User's manual,''
Phys.\ Rept.\  {\bf 388}, 279 (2003).
[hep-th/0306138].
}

\lref\CamporesiWN{
  R.~Camporesi and A.~Higuchi,
  ``Stress energy tensors in anti-de Sitter space-time,''
Phys.\ Rev.\ D {\bf 45}, 3591 (1992).
}

\lref\CamporesiGA{
  R.~Camporesi and A.~Higuchi,
  ``Spectral functions and zeta functions in hyperbolic spaces,''
J.\ Math.\ Phys.\  {\bf 35}, 4217 (1994).
}

\lref\BanerjeeQC{
  S.~Banerjee, R.~K.~Gupta and A.~Sen,
  ``Logarithmic Corrections to Extremal Black Hole Entropy from Quantum Entropy Function,''
JHEP {\bf 1103}, 147 (2011).
[arXiv:1005.3044 [hep-th]].
}

\lref\BanerjeeJP{
  S.~Banerjee, R.~K.~Gupta, I.~Mandal and A.~Sen,
  ``Logarithmic Corrections to N=4 and N=8 Black Hole Entropy: A One Loop Test of Quantum Gravity,''
JHEP {\bf 1111}, 143 (2011).
[arXiv:1106.0080 [hep-th]].
}

\lref\SenBA{
  A.~Sen,
  ``Logarithmic Corrections to N=2 Black Hole Entropy: An Infrared Window into the Microstates,''
[arXiv:1108.3842 [hep-th]].
}

\lref\SenAJA{
  A.~Sen,
  ``Microscopic and Macroscopic Entropy of Extremal Black Holes in String Theory,''
[arXiv:1402.0109 [hep-th]].
}

\lref\NicolaiTD{
  H.~Nicolai and P.~K.~Townsend,
  ``N=3 Supersymmetry Multiplets with Vanishing Trace Anomaly: Building Blocks of the $N\ge3$ Supergravities,''
Phys.\ Lett.\ B {\bf 98}, 257 (1981).
}

\lref\FerraraIH{
  S.~Ferrara, R.~Kallosh and A.~Strominger,
  ``N=2 extremal black holes,''
Phys.\ Rev.\ D {\bf 52}, 5412 (1995).
[hep-th/9508072].
}

\lref\followup{F.~Larsen and P.~Lisb\~{a}o, to appear.}

\lref\HawkingJA{
  S.~W.~Hawking,
  ``Zeta Function Regularization of Path Integrals in Curved Space-Time,''
Commun.\ Math.\ Phys.\  {\bf 55}, 133 (1977)..
}

\lref\ChristensenMD{
  S.~M.~Christensen and M.~J.~Duff,
  ``New Gravitational Index Theorems and Supertheorems,''
Nucl.\ Phys.\ B {\bf 154}, 301 (1979)..
}

\lref\GuptaHXA{
  R.~K.~Gupta, S.~Lal and S.~Thakur,
  ``Logarithmic Corrections to Extremal Black Hole Entropy in N = 2, 4 and 8 Supergravity,''
[arXiv:1402.2441 [hep-th]].
}

\lref\GuptaSVA{
  R.~K.~Gupta, S.~Lal and S.~Thakur,
  ``Heat Kernels on the $AdS_2$ cone and Logarithmic Corrections to Extremal Black Hole Entropy,''
JHEP {\bf 1403}, 043 (2014).
[arXiv:1311.6286 [hep-th]].
}

\lref\ArosTAA{
  R.~Aros, D.~E.~Diaz and A.~Montecinos,
  ``On Wald entropy of black holes: logarithmic corrections and trace anomaly,''
[arXiv:1305.4647 [gr-qc]].
}

\lref\PourdarvishGFA{
  A.~Pourdarvish, J.~Sadeghi, H.~Farahani and B.~Pourhassan,
  ``Thermodynamics and Statistics of Goedel Black Hole with Logarithmic Correction,''
Int.\ J.\ Theor.\ Phys.\  {\bf 52}, 3560 (2013).
}

\lref\BhattacharyyaWZ{
  S.~Bhattacharyya, B.~Panda and A.~Sen,
  ``Heat Kernel Expansion and Extremal Kerr-Newmann Black Hole Entropy in Einstein-Maxwell Theory,''
JHEP {\bf 1208}, 084 (2012).
[arXiv:1204.4061 [hep-th]].
}

\lref\DasIC{
  S.~Das, P.~Majumdar and R.~K.~Bhaduri,
  ``General logarithmic corrections to black hole entropy,''
Class.\ Quant.\ Grav.\  {\bf 19}, 2355 (2002).
[hep-th/0111001].
}

\lref\SenDW{
  A.~Sen,
  ``Logarithmic Corrections to Schwarzschild and Other Non-extremal Black Hole Entropy in Different Dimensions,''
JHEP {\bf 1304}, 156 (2013).
[arXiv:1205.0971 [hep-th]].
}

\lref\SenCJ{
  A.~Sen,
  ``Logarithmic Corrections to Rotating Extremal Black Hole Entropy in Four and Five Dimensions,''
Gen.\ Rel.\ Grav.\  {\bf 44}, 1947 (2012).
[arXiv:1109.3706 [hep-th]].
}

\lref\WaldNT{
  R.~M.~Wald,
  ``Black hole entropy is the Noether charge,''
Phys.\ Rev.\ D {\bf 48}, 3427 (1993).
[gr-qc/9307038].
}

\lref\JacobsonVJ{
  T.~Jacobson, G.~Kang and R.~C.~Myers,
  ``On black hole entropy,''
Phys.\ Rev.\ D {\bf 49}, 6587 (1994).
[gr-qc/9312023].
}

\lref\CamporesiNW{
  R.~Camporesi,
  ``zeta function regularization of one loop effective potentials in anti-de Sitter space-time,''
Phys.\ Rev.\ D {\bf 43}, 3958 (1991)..
}

\lref\SenVM{
  A.~Sen,
  ``Quantum Entropy Function from AdS(2)/CFT(1) Correspondence,''
Int.\ J.\ Mod.\ Phys.\ A {\bf 24}, 4225 (2009).
[arXiv:0809.3304 [hep-th]].
}

\lref\StromingerKF{
  A.~Strominger,
Phys.\ Lett.\ B {\bf 383}, 39 (1996).
[hep-th/9602111].
}

\lref\GopakumarQS{
  R.~Gopakumar, R.~K.~Gupta and S.~Lal,
JHEP {\bf 1111}, 010 (2011).
[arXiv:1103.3627 [hep-th]].
}

\Title{\vbox{\baselineskip12pt\
}}
{\vbox{\centerline{Logarithmic Corrections to ${\cal N}\geq 2$ Black Hole Entropy}}}
\centerline{
Cynthia Keeler\foot{keelerc@umich.edu},  Finn
Larsen\foot{larsenf@umich.edu}, and Pedro Lisb\~{a}o\foot{plisbao@umich.edu}
}
\bigskip
\centerline{\it{Department of Physics
and Michigan Center for Theoretical Physics,}}
\centerline{\it{University of Michigan, Ann Arbor, MI 48109-1120, USA.}}

\baselineskip15pt

\vskip .3in

\centerline{\bf Abstract}
We revisit the computation of logarithmic corrections to black holes with ${\cal N}\geq 2$ supersymmetry. We employ an on-shell
method that takes advantage of the symmetries in the AdS$_2\times S^2$ near horizon geometry. For bulk modes interactions
are incorporated through the spectrum of chiral primaries that we derive afresh. The spectrum of boundary states is computed explicitly by analyzing gauge variations. Elementary heat kernels in 4D and 2D then give the logarithmic corrections to the black hole entropy. Our computation represents a streamlined and simplified derivation that agrees with the results recently found by A. Sen.

\Date{April, 2014}
\baselineskip14pt

\newsec{Introduction}

The microscopic understanding of black hole quantum states gives counting formulae that encode numerous corrections to the standard Bekenstein-Hawking area law. Among such corrections the logarithmic ones have particular significance because these are independently computable from the effective low energy theory in the vicinity of the black holes \refs{\SenVM, \SenBA, \SenAJA}. The logarithmic corrections therefore allow a sensitive check on any proposed microscopic model. Conversely, in cases where no microsopic model is available the logarithmic corrections provide a robust clue that may lead to the construction of such a model \refs{\SenCJ,\SenDW}.

The computation of logarithmic corrections from the low energy theory is straightforward in principle \HawkingJA: determine the quadratic fluctuations around the black hole background and then compute the resulting functional determinant using standard techniques. However, in practice these steps can be quite laborious. The theories of interest in string theory generally have elaborate matter content that results in many distinct contributions to quantum corrections. Gauge symmetries (including diffeomorphism invariance) further complicate the situation by introducing ghost sectors that can be quite nontrivial. The logarithmic corrections to black holes were developed in many
recent works including \refs{\DasIC,\BhattacharyyaWZ,\PourdarvishGFA,\ArosTAA,\GuptaSVA}.

The goal of this paper is to present a simplified computation of logarithmic corrections to the black hole entropy. The streamlined procedure we present promotes transparency and makes it realistic to address more complicated settings. In this paper we limit ourselves to BPS black holes which have AdS$_2\times S^2$ near horizon geometry. In this context important aspects of our strategy are these:

\itemitem{$\bullet$}
{\it The Spectrum of Chiral Primaries:} a large number of interactions between different fields generally leads to unwieldy matrices at intermediate stages of the computation. We diagonalize the interactions by first computing the spectrum of chiral primaries. This spectrum encodes all information about the interactions that is needed.
\itemitem{\phantom{$\bullet$}}\indent
In order to highlight the origin of these simplifications in symmetry principles we give a self-contained derivation of the black hole spectrum. Our method is indirect but it is efficient and new to this context. Further, our independent computation of the spectrum identifies several details that have previously been overlooked.

\itemitem{$\bullet$}
{\it Simplified Functional Determinants:} we reduce the field content of the 4D theory to a set of fields on AdS$_2$ and its boundary. The only functional determinants we need are those for massless scalars and fermions in AdS$_2$. The additional data that is special to each field we consider is encoded in a discrete sum over masses. This organization of the computation represents a simplification because it does not require measures and contours for continuous complex eigenvalues. We also do not need explicit wave functions.

\itemitem{$\bullet$}
{\it Gauge-fixing and Ghosts:} we compute quantum corrections by summing over contributions from physical fields only. The unphysical sector comprising pure gauge modes, longitudinal modes, and ghosts ultimately cancel in the physical quantities of interest. We use an on-shell method where these quantities are not needed in intermediate stages of the computation.

\itemitem{$\bullet$}
{\it Boundary Modes:} gauge symmetries (including supersymmetry and diffeomorphism invariance) give rise to physical modes that localize on the boundary. We determine the quantum numbers of these modes by analyzing the action of the relevant symmetry.  Their contribution is then computed by treating them as 2D fields on the $S^2$.

\medskip
The physical modes that contribute to the one-loop functional determinant are the 4D bulk modes, the 2D boundary modes, and the 0D zero-modes. Adding the contributions together our final result for logarithmic corrections to extremal black hole entropy in theories with ${\cal N}\geq 2$ SUSY
becomes
\eqn\aa{
S = {1\over 4}A_H+ {1\over 12} \left[ 23 - 11 ({\cal N}-2) - n_V + n_H\right] \log A_H~.
}
This final result agrees perfectly with those
reported by A. Sen and collaborators \refs{\BanerjeeQC,\BanerjeeJP,\SenBA}.
Some important special cases of the formula:

\itemitem{$\bullet$}
The ${\cal N}=4$ theory. Such theories have $n_V=n_H+1$ because one ${\cal N}=2$ vector is part of the ${\cal N}=4$ supergravity multiplet while each ${\cal N}=4$ matter multiplet is composed of one ${\cal N}=2$ vector and one ${\cal N}=2$ hyper. In this case the logarithmic correction vanishes independently of the number of ${\cal N}=4$ matter multiplets.

\itemitem{$\bullet$}
The ${\cal N}=6$ theory: $n_V=7$ and $n_H=4$ so that the logarithmic correction is $\delta S = - 2 \log A_H$.

\itemitem{$\bullet$}
The ${\cal N}=8$ theory: $n_V=15$ and $n_H=10$ so
$
\delta S = - 4 \log A_H~.
$
\medskip

We evaluate the functional determinants using heat kernel techniques. In 4D the leading term in the heat kernel is a double pole. These double poles cancel in each ${\cal N}=2$ multiplet by itself. This corresponds to vanishing cosmological constant in 4D and is due to the degeneracy of bosons and fermions in the on-shell SUSY multiplets.

The simple pole in the heat kernel receives contributions from the 2D boundary modes that are non-trivial since there is not the same number of bosonic and fermionic symmetries. It also receives a contribution from mixing between the bulk modes. It is a consistency check on our computations that the sum of these terms vanish for any theory with at least ${\cal N}=4$ \ChristensenMD. For the more general theories we consider the coefficient of the pole in the heat kernel is non-trivial. This part of our result can be interpreted as the renormalization of the gravitational coupling constant.

The logarithmic corrections to the black hole entropy are encoded in the constant term of the heat kernel so contributions from both bulk modes and boundary modes must be computed with sufficient precision that the constant is determined. Additionally, there are contributions from zero-modes.

The indirect methods we pursue in this paper stress the origin of particle spectra in symmetry but at times they leave room for suspicion. In a companion paper we will present the explicit mode expansions that underpin the physical spectrum \followup.

This paper is organized as follows. In section 2 we determine the spectrum of chiral primaries using an indirect argument that exploits symmetries. We resolve a discrepancy with results reported in the literature. In section 3 we review the simple heat kernels we need. We provide a self-contained presentation in order to highlight the complete absence of advanced techniques. In section 4 we apply the heat kernels to the physical spectrum determined in section 2. We thus compute the contribution to the heat kernel from all bulk modes.
In section 5 we discuss gauge symmetries and use them to determine the spectrum of boundary modes. This yields an additional contribution to the heat kernel. In section 6 we briefly review the correction to the heat kernel due to zero-modes on the boundary. Finally, in section 7 we add the various contibutions to the heat kernel and we discuss the relation to trace anomalies. This gives the logarithmic correction to the black hole entropy \aa.

\newsec{Classical Modes}
The {\it spectrum} of the black hole is the set of quantum numbers for fluctuations around the black hole background. In this section we use symmetry principles to determine the BPS part of the spectrum.

We consider a 4D theory with (at least) ${\cal N}=2$ SUSY. We further focus on the near horizon region of black holes that preserve at least some of the supersymmetry. This geometry always takes the form AdS$_2\times S^2$. The attractor mechanism ensures that gravity and the graviphoton are the only fields turned on in the near horizon geometry of the black hole \FerraraIH.

Fields in the AdS$_2\times S^2$ background are classified by the quantum numbers of the $SL(2)\times SU(2)$ isometries. We are particularly interested in the lowest weight representations which we denote by $(h,j)$. Here $h$ is the lowest eigenvalue of the $L_0$ generator of $SL(2)$ and $j$ refers to the $SU(2)$ representation. The $(h,j)$ representation thus has degeneracy $(2j+1)$ from its $SU(2)$ representation and also an infinite tower of states with $L_0$ values $h, h+1, h+2,\ldots$. The BPS spectrum of the black hole is a list of the $(h,j)$ that are realized by fluctuations in the background.

The massless field content of a general theory with ${\cal N}\geq 2$ SUSY can be decomposed into a set of ${\cal N}=2$ multiplets:

\itemitem{$\bullet$}
A  supergravity multiplet.

\itemitem{$\bullet$}
${\cal N}-2$ (massive) gravitino multiplets (because two of the ${\cal N}$ gravitinos are in the ${\cal N}=2$ supergravity multiplet).

\itemitem{$\bullet$}
$n_V$ vector multiplets.

\itemitem{$\bullet$}
$n_H$ hyper multiplets.

\subsec{Determination of BPS Spectra}

It is useful to organize the particle content of ${\cal N}=2$ multiplets according to their {\it helicity content}. Suppose that the maximum helicity state in a given ${\cal N}=2$ multiplet is $\lambda$. Upon action with one of the two SUSY generators we then find two states with helicity $\lambda-{1\over 2}$ and, upon action with both of them, we find a single state with helicity $\lambda-1$. This universal structure gives the helicity content of each ${\cal N}=2$ multiplet:
\eqn\ba{\eqalign{
{\rm Supergravity~multiplet:} &~ \lambda=\pm 2, \pm {3\over 2}\times 2, \pm 1~,\cr
{\rm Gravitino~multiplet:} &~ \lambda=\pm {3\over 2}, \pm 1\times 2, \pm {1\over 2}~,\cr
{\rm Vector~multiplet:} &~ \lambda=\pm 1, \pm {1\over 2}\times 2, 0\times 2~,\cr
{\rm Hyper~multiplet:} &~ \lambda=\pm {1\over 2}\times 2, 0\times 4~.
}}
\noindent
The notation $\times 2$ indicates a multiplicity of $2$.
In the first three kinds of multiplets we included the CPT conjugate states with negative helicity as one must in field theory realizations. The hypermultiplet was automatically CPT invariant but we double its field content anyway. With this convention the hypermultiplet is a ``full" hyper with $4$ real scalars and two Weyl spinors.

The field equations for quadratic fluctuations are linear. Moreover, we can introduce global flavor symmetries unique to each type of ${\cal N}=2$ supermultiplet and this ensures that there is no mixing between different types of ${\cal N}=2$ supermultiplets. We can therefore consider the supergravity multiplet, the (massive) gravitino multiplets, the vector multiplets, and the hyper multiplets independently.

The expansion of four-dimensional fields in partial waves on $S^2$ gives an effective 2D theory on AdS$_2$. The $SU(2)$ representations that appear are determined by the general rules that govern Kaluza-Klein reduction on homogeneous spaces \SalamXD. In the case of the coset $S^2=SU(2)/U(1)$ the quantum number under $U(1)$ can be identified with the helicity $\lambda$ and the $SU(2)$ representations that appear in the reduction are precisely those where $\lambda$ appears in the decomposition of $SU(2)$ with respect to $U(1)$. Thus the allowed angular momentum quantum numbers for a helicity mode $\lambda$ are $j=|\lambda|, |\lambda|+1,\ldots$. Starting from the helicity content of the fields \ba\ we can therefore present the $SU(2)$ content in terms of towers:
\eqn\bb{\eqalign{
{\rm Supergravity~multiplet:} &~ j=(k+2)\times 2, (k+{3\over 2})\times 4, (k+1)\times 2~,\cr
{\rm Gravitino~multiplet:} &~ j=(k+{3\over 2})\times 2, (k+1)\times 4, (k+{1\over 2})\times 2~, \cr
{\rm Vector~multiplet:} &~ j=(k+1)\times 2, (k+{1\over 2})\times 4, k\times 2~, \cr
{\rm Hyper~multiplet:} &~ j=(k+{1\over 2})\times 4, k\times 4~,
}}
with $k=0,1,\ldots$.

\medskip
The BPS spectrum of the black hole amounts to the specification of the value of the AdS$_2$ energy $h$ for each of these $SU(2)$ multiplets. These energies depend on couplings between the fields. The simplification captured by the enumeration in \bb\ is that these couplings respect the partial wave expansion: only fields with the same $j$ can mix.

The actual value of the AdS$_2$ energy $h$ is determined by supersymmetry as follows. The AdS$_2\times S^2$ geometry preserves the supergroup $SU(2|1,1)$. This supergroup has $8$ SUSY charges, the same as the number in ${\cal N}=2$ SUSY in four dimensions. These generators can be represented in terms of two component spinors ${\cal Q}^A$ ($A=1,2$) and their conjugates. The corresponding charges all have quantum numbers $h=1/2$ and $j=1/2$. They  transform as doublets of the global $SU(2)$ symmetry acting on the $A=1,2$ index. We will suppress reference to this global $SU(2)$ in the following in order to avoid confusion with the $SU(2)$ rotation group.  Since SUSY is preserved by the background, fluctuating fields must organize themselves into supermultiplets after the mixing is taken into account. Starting from a lowest weight state $(h,j)$ a supermultiplet is obtained by acting with the supercharges that function as creation operators.

The fields we consider will all be in chiral multiplets of the form
\eqn\bc{
(k,k), 2(k+{1\over 2},k-{1\over 2}), (k+1,k-1)~,
}
with the possible values of $k={1\over 2}, 1, {3\over 2}, \ldots$. In the special case where $k={1\over 2}$ the $SU(2)$ quantum number $j=-{1\over 2}$ of the final term in \bc\ should be interpreted as an empty
representation.

The chiral multiplets \bc\ are {\it short} multiplets. They are special in two (related) ways: the lowest weight state has $h=j$ and also the supercharges always act in a manner that ${\it lowers}$ the spin. A generic long representation would have four active supercharges so that the span of spins in a single multiplet would be two. Such representations are therefore too large for our purpose.

There is a unique way to organize the fields with $SU(2)$ content \bb\ into chiral multiplets of the form \bc.
This gives the list of fields
\eqn\bd{\eqalign{
{\rm Supergravity ~multiplet:} &~2[(k+2,k+2)~, 2(k+{5\over 2},k+{3\over 2})~, (k+3,k+1)]~,\cr
{\rm Gravitino ~multiplet:} &~2[(k+{3\over 2},k+{3\over 2})~, 2(k+2,k+1)~, (k+{5\over 2},k+{1\over 2} )]~,\cr
{\rm Vector ~multiplet:} &~2[(k+1,k+1)~, 2(k+{3\over 2},k+{1\over 2})~, (k+2,k)]~\cr
{\rm Hyper ~multiplet:} &~2[(k+{1\over 2},k+{1\over 2})~, 2(k+1,k)~, (k+{3\over 2},k-{1\over 2})]~.
}}
As before $k=0,1\ldots$. This is the complete spectrum of the black hole. In particular the spectrum is determined entirely by symmetries.

\subsec{Explicit Computations}
The determination of the on-shell spectrum using symmetry constraints illuminates its group theory origin. However, the indirect nature of the method may leave some conceptual unease. It is therefore worthwhile to consider an alternative, the explicit diagonalization of the action expanded to quadratic order. This approach was carried out over a decade ago for the case of pure ${\cal N}=2$ SUGRA \MichelsonKN\ and for the maximally supersymmetric theory with ${\cal N}=8$ SUSY \refs{\CorleyUZ,\LeeYU}. Combination of the final tables in these references yields towers of multiplets that can be compared with our results \bd\ that apply to the slightly more general case where ${\cal N}=2$ SUGRA is coupled to ${\cal N}-2$ (massive) gravitini multiplets, $n_V$ vector multiplets, and $n_H$ hypermultiplets. The results in the references agree precisely with \bd\ {\it with one exception}: all previous works report an additional chiral multiplet. In our notation the additional states that were reported correspond to the extension of one of the two supergravity multiplet towers in \bd\ to include the mode $k=-1$. Thus the primary states reported
in \refs{\MichelsonKN,\CorleyUZ,\LeeYU} but absent from our analysis are
\eqn\ca{
(1,1)~, 2({3\over 2}, {1\over 2}), (2,0)~.
}
It is instructive to find the origin of this discrepancy.

As a starting point for this specific purpose it is sufficient to consider 4D Einstein gravity coupled to a $U(1)$ gauge field
\eqn\cba{
{\cal L}_4 = {1\over 16\pi G} \left[ {\cal R}^{(4)} - {1\over 4} F_{IJ}F^{IJ}\right]~.
}
We use 4D indices $I,J,\ldots$, AdS$_2$ indices $\mu,\nu,\ldots$, and $S^2$ indices $\alpha,\beta,\ldots$. One solution to this theory is the AdS$_2\times S^2$ geometry supported by the magnetic monopole $F_{\alpha\beta} = 2\epsilon_{\alpha\beta}$. With this normalization the AdS$_2$ and $S^2$ radii are both ``1". The Freund-Rubin reduction on $S^2$ is realized by the 4D geometry:
\eqn\cb{ ds^2 = g_{\mu\nu} dx^\mu dx^\nu + X d\Omega^2_2~,
}
where $g_{\mu\nu}$ and $X$ are arbitrary functions of the 2D coordinates $x^\mu$, $\mu=1,2$. The effective 2D Lagrangian becomes
\eqn\cd{
{\cal L}_2 = {1\over 4G} \left[ X{\cal R}^{(2)} + 2 - {2\over X} + {(\nabla X)^2\over 2X} \right]~.
}
The equations of motion are obtained upon variation of ${\cal L}_2$ by the scalar $X$
\eqn\ce{
{\cal R}^{(2)} + {2\over X^2} + {(\nabla X)^2\over 2X^2} - {1\over X}\nabla^2 X =0~,
}
and by the metric $g_{\mu\nu}$
\eqn\cea{
X ({\cal R}^{(2)}_{\mu\nu} - {1\over 2} g_{\mu\nu}{\cal R}^{(2)}) + {1\over 2X} [\nabla_\mu X\nabla_\nu X
- {1\over 2} g_{\mu\nu} (\nabla X)^2] - g_{\mu\nu} (1-{1\over X}) + g_{\mu\nu} \nabla^2 X - \nabla_\mu\nabla_\nu X  =0~.
}
Recall that the Riemann tensor has just a single component in 2D so after contractions ${\cal R}^{(2)}_{\mu\nu} = {1\over 2} g_{\mu\nu}{\cal R}^{(2)}$ identically for any 2D geometry, not just for symmetric geometries. The first term in \cea\ therefore vanishes identically. We write this term temporarily because it reminds us that \cea\ is the Einstein equation while \ce\ is the equation of motion for the 2D matter field $X$. As a check on \ce\ and \cea\ note that the AdS$_2$ geometry satisfies these equations with ${\cal R}^{(2)}=-2$ and $X=1$. This corresponds to AdS$_2$ and $S^2$ radii equal to ``1".

The Einstein equation \cea\ decomposes into the trace
\eqn\cf{
\nabla^2 X = 2(1-{1\over X})~,
}
and (upon use of \cf) the traceless equation
\eqn\cfa{
(\nabla_\mu\nabla_\nu-{1\over 2}g_{\mu\nu}\nabla^2)\sqrt{X}=0~.
}
Taking \cf\ in isolation we find that small variations $\delta X$ around the background $X=1$ satisfy a Klein-Gordon equation with $m^2=2$. In AdS$_2$ scalar excitations with this mass have conformal weight $h=2$. The excitations described by the Freund-Rubin compactification \cb\ are spherically symmetric (j=0) so this mode would have quantum numbers $(h,j)=(2,0)$. Comparison with \ca\ shows that this is exactly the mode that the explicit analyses recognize as physical but our indirect analysis does not. We will show that the discrepancy is due to the constraints expressed by \cfa.

For perspective on the discrepancy recall the elementary counting of degrees of freedom. Perturbative 2D gravity is described by the symmetric tensor $\delta g_{\mu\nu}= h_{\mu\nu}$ with $3$ components. Diffeomorphisms $\delta_\xi h_{\mu\nu} = \partial_\mu \xi_\nu + \partial_\nu \xi_\mu$ impose equivalences that render two components of $h_{\mu\nu}$ redundant. The equations of motion resulting from variations of those two components further impose two constraints so the net number of degrees of freedom in pure 2D gravity is -1. This awkward counting is special to 2D where it is indeed well known for theories such as dilaton gravity (see eg \GrumillerNM). It implies that the {\it combination} of 2D gravity (described by $h_{\mu\nu}$) {\it and} a scalar field (in the present context the 2D scalar field $X$) will have no degrees of freedom.

There are several known exceptions to this simple type of counting: there may be important quantum effects (captured by a class of matrix models) or there may be classical degrees of freedom in less than 2D. In the present context there are indeed 1D boundary states but they should not be confused with bulk degrees of freedom which is where we differ from previously reported results.

To make the general discussion on the counting of degrees of freedom more explicit we fix the gauge $g_{zz}=1, g_{zt}=0$ and so consider the 2D geometry in the form
\eqn\cg{
ds^2 =  - e^{2\rho} dt^2 + dz^2~,
}
where $\rho=\rho(t,z)$ is an arbitrary function. In this gauge we can represent the background AdS$_2$ as either just the Poincar\'{e} patch (with $e^{2\rho_0}=e^{2z}$) or global AdS$_2$ (with $e^{2\rho_0}=\cosh^2 z$) or as an AdS$_2$ black hole (with $e^{2\rho_0}=\sinh^2z$). For any of these backgrounds the $zz$ and $zt$-components of \cfa\ give
\eqn\ci{\eqalign{
(\partial^2_z - 1 ) \delta X & =0~,\cr
\partial_z (e^{-\rho_0}\partial_t \delta X) &=0~,
}}
after linearization.
The first equation was simplified using \cf. These equations are {\it constraints} on fluctuations $\delta X$. If $\delta X$ were a propagating field, we would be able to specify $\delta X$ and its time derivative $\partial_t \delta X$ for all $z$ at an initial time and then use the equations of motion to find $\delta X$ at later times. The constraints \ci\ show that this is impossible: once we have given $\delta X$ and $\partial_t \delta X$ for large $z$, initial conditions are specified for all $z$. Thus $\delta X$ is in fact a boundary degree of freedom.

We have not yet analyzed the equation of motion \ce\ which relates the curvature ${\cal R}^{(2)}$ to the scalar field $X$. The Ricci curvature of \cg\ is
\eqn\cj{
{\cal R}^{(2)} = -2e^{-\rho} \partial^2_z e^\rho~,
}
so \ce\ can be recast as
\eqn\ck{
2\delta X = \nabla^2\delta X = - {1\over 2}\delta {\cal R}^{(2)} = e^{-\rho_0} (\partial^2_z-1) \delta\rho ~.
}
This demonstrates that perturbations $\delta\rho$ with $\partial_z^2\delta\rho=1$ are independent degrees of freedom.

In summary, in this subsection we analyzed the spherically symmetric sector of gravity comprising the 2D metric $h_{\mu\nu}$ and the scalar field $X$ encoding the size of the $S^2$. We find that after taking gauge fixing and constraints into account the bulk theory has no physical states but two boundary degrees of freedom remain.

\subsec{Boundary Modes}
The table \bd\ enumerates all {\it bulk} modes of the black holes. In addition to these modes there are {\it boundary} modes. The boundary modes are closely associated with gauge symmetries of the theory. Each component of a gauge symmetry allows the removal of one component field. Additionally, the equation of motion for the component thus removed ceases to be dynamical: it becomes a {\it constraint}. As discussed in the previous subsection, constraints limit the dynamics of the theory by restricting the independent initial data. In the context of AdS$_2$ each constraint gives rise to one boundary mode.

We first consider the supergravity multiplet. The perturbation $h_{IJ}$ of the 4D metric has $10$ components. Diffeomorphisms $\delta_\xi h_{IJ} = \partial_I\xi_J +\partial_J\xi_I$ are generated by the vector field $\xi_I$ with $4$ components. Thus the graviton has $6$ components subject to $4$ constraints. This yields a net of $2$ physical degrees of freedom in bulk, as it should. But in addition the boundary data on the $4$ constraints give rise to $4$ boundary degrees of freedom. These boundary degrees of freedom have the quantum numbers of the diffeomorphism generator $\xi_I$. In particular, they have helicity content $\lambda=\pm 1, 0,0$.

A chiral gravitino $\Psi_I$ has $6$ components after the Rarita-Schwinger constraint $\gamma^I\Psi_I=0$ is taken into account. After gauge fixing of local supersymmetry (generated by a chiral spinor with two components) it has $4$ components subject to two constraints. This yields a net of two physical degrees of freedom but also two boundary degrees of freedom.

Finally, the graviphoton ${\cal A}_I$ has four components. The $U(1)$ gauge symmetry removes one component so three components remain which are subject to one constraint. This gives two physical components for the graviton in bulk but also a single boundary degree of freedom.

Proceeding similarly for the (massive) gravitino multiplet and the vector multiplet, the helicity content of all physical boundary modes becomes
\eqn\da{\eqalign{
{\rm Supergravity~multiplet:} & ~\lambda = \pm 1, \pm{1\over 2} \times 2, 0\times 3~,\cr
{\rm Gravitino~multiplet:} & ~\lambda = \pm{1\over 2}, 0\times 2~,\cr
{\rm Vector~multiplet:} & ~\lambda = 0~.
}}
The hyper multiplet is not mentioned since it does not have any gauge degrees of freedom and therefore no boundary states.
The helicity content \da\ in turn determines the $SU(2)$ content of the boundary modes as
\eqn\db{\eqalign{
{\rm Supergravity~multiplet:} &~ j=(k+1)\times 2, (k+{1\over 2})\times 4, k\times 3~,\cr
{\rm Gravitino~multiplet:} &~ j=(k+{1\over 2})\times 2, k\times 2~, \cr
{\rm Vector~multiplet:} &~ j=k~, \cr
}}
with $k=0,1,\ldots$.

Our discussion of boundary states here focuses on gauge invariance. As such it is based on the off-shell (un-physical) components of the various fields. The on-shell supersymmetry realized by the fields we consider does not extend a simple way to these off-shell degrees of freedom. In the absence of further data it is therefore not possible to compute the conformal weights of these fields from superconformal invariance alone.

Two of the three $j=0$ fields in the supergravity multiplet are the $\delta\rho$ and $\delta X$ discussed explicitly in the previous subsection. Similar computations for the remaining fields determine the full spectrum of boundary states \followup. In section 5 we determine spectrum of boundary states by exploiting symmetries.

\newsec{The Heat Kernel Expansion: Elementary Examples}
This section reviews the basics of the heat kernel method \refs{\HawkingJA, \VassilevichXT, \GopakumarQS}. We introduce notation and also give elementary evaluations of the key examples that later will be generalized.

\subsec{Functional Determinants and the Heat Kernel}
One loop quantum corrections are encoded in Euclidean path integrals taking a Gaussian form which we present schematically as
\eqn\ka{
e^{-W} = \int {\cal D}\phi ~e^{-\phi\Lambda\phi} = {1\over\sqrt{{\rm det}\Lambda}}~.
}
The kinetic operator generally includes a mass term $\Lambda=-\Delta+m^2$. We suppress the indices on $\phi$ that enumerate components of the field such as those that incorporate Lorentz structure.

After UV regulation the effective action $W$ becomes
\eqn\kb{
W = {1\over 2} \ln{\rm det} \Lambda= {1\over 2}\sum_i \ln\lambda_i
= -{1\over 2}\int^\infty_{\epsilon^2} ds {D(s)\over s}~,
}
where $\{\lambda_i\}$ are the eigenvalues of $\Lambda$ and the heat kernel
\eqn\kc{
D(s) = {\rm Tr} ~e^{-s\Lambda} = \sum_i e^{-s\lambda_i}~.
}
We use a notation where the eigenvalues $\lambda_i$ are assumed discrete even though in practice they may be continuous. Also, in cases where the fields are fermionic the determinant in \ka\ should be in the numerator instead and then the contribution to the effective action \kb\ will enter with the opposite sign.

The heat kernel terminology arises because it is often useful to express $D(s)$ as
\eqn\kd{
D(s) = \int d^D x ~K(x,x;s)~,
}
where the Green's function satisfies the heat equation
\eqn\ke{
\left( \partial_s + \Lambda_x\right) K(x,x';s)=0~,
}
with the boundary condition $K(x,x';s)=\delta(x-x')$ at $s=0$.
The Green's function can be expanded on a complete basis as
\eqn\kf{
K(x,x';s) = \sum_i e^{-s\lambda_i} f_i(x) f_i^*(x')~,
}
where $\{f_i\}$ are the normalized eigenfunctions of $\Lambda$ with eigenvalues $\{\lambda_i\}$.
Inserting this expansion in \kd\ and using the normalization condition we do indeed recover \kc.

As an example, in flat space with $D$ Euclidean dimensions the eigenfunctions of the kinetic operator are plane waves $e^{ikx}$ and the eigenvalues are $k^2+m^2$. The expression \kf\ becomes a Gaussian integral which upon integration gives the Green's function
\eqn\kg{
K_{\rm flat} (x,x';s) = \left( {1\over 4\pi s}\right)^{D\over 2} e^{-{1\over 4s} (x-x')^2- m^2 s}~.
}
Inserting this expression in \kd\ we find the heat kernel for a massless scalar field
\eqn\kh{
D_{\rm flat}(s) = \left( {1\over 4\pi s}\right)^{D\over 2}  {\rm Vol}~.
}
This expression gives the leading asymptotic behavior for small $s$ (small distance) in any geometry. A standard approach to curved space examples is to correct the flat space result \kh\ perturbatively (see {\it eg.} \VassilevichXT). This gives an expansion in small $s$ with coefficients that are scalars formed from the curvature. For example, for a minimally coupled scalar field
\eqn\kha{
K^s(s) = \left({1\over 4\pi s}\right)^{D\over 2} \left[ 1 + {s\over 6}{\cal R}  +{s^2\over 360} (5{\cal R}^2 - 2R_{IJ}R^{IJ}
+ 2R_{IJKL}R^{IJKL}) +\ldots \right]~.
}
Similar expansions apply to other fields.

In our computations we will actually {\it not} employ the heat equation \ke\ and, related to that, we will avoid the explicit eigenfunctions. Instead we will compute $D(s)$ directly from \kc\ by explicit summation over eigenvalues. In the homogeneous spaces we focus on the corresponding heat kernel density is then given by
\eqn\ki{
K(s) = {1\over {\rm Vol}}D(s)~.
}
For a sphere $S^2$ with radius $a$ the volume is simply ${\rm Vol}_S=4\pi a^2$. For AdS$_2$ the volume diverges but it can be regulated near the boundary
\eqn\kig{
{\rm Vol}_A = 2\pi a^2 \int_0^{\rho_{\rm max}} d\rho \sinh\rho   = 2\pi a^2( \cosh\rho_{\rm max} - 1)~.
}
In the context of AdS/CFT it is often appropriate to remove the $\cosh\rho_{\rm max}$ by adding terms that are intrinsic and local on the boundary. This gives ${\rm Vol}_A =  (-2\pi a^2)$ for the renormalized volume of AdS$_2$. We do not use this value since a positive
volume makes it easier to track signs for fermions and bosons. The dependence of the actual (regulated) volume \kig\ on a cut-off will anyway cancel in physical results so we can effectively take ${\rm Vol}_A =  +2\pi a^2$ when an explicit volume is needed.

Although our strategy is to compute $D(s)$ using the sum \kc\ we will quote results in terms of $K(s)$ using the relation \ki. This practice will facilitate comparison with the literature.

\subsec{The Scalar on $S^2$}
The heat kernel on the two-sphere $S^2$ is of special importance to us since it will serve as the building block for all our computations.

The determination of this heat kernel is particularly simple because the eigenvalue problem of the Laplacian on $S^2$ has been studied by all physics students since their first course in quantum mechanics. The possible eigenvalues of $-\nabla^2$ are $l(l+1)$ with each value of the orbital angular momentum $l=0,1,\ldots$ appearing with degeneracy $2l+1$ corresponding to the possible azimuthal quantum numbers $m=-l,\ldots, l$. The corresponding eigenfunctions are the spherical harmonics $Y_{lm}$. These basic facts immediately give the heat kernel (density) for a minimally coupled scalar field on $S^2$:
\eqn\ha{\eqalign{
K^s_{S}(s) &= {1\over 4\pi a^2} \sum_{k=0}^\infty ~e^{-sk(k+1)} (2k+1) ~.
}}
We can expand for small $s$ using the Euler-MacLaurin formula in the form simplified for functions with $f^{(n)}(\infty) =0$:
\eqn\hb{\eqalign{
\sum^\infty_{k=0} f(k) & = \int^\infty_{0} dk f(k)  + {1\over 2}(f(0) + f(\infty) ) + \sum_{n=1}^\infty {B_{2n}\over (2n)!}
\left( f^{(2n-1)} (\infty)  - f^{(2n-1)} (0) \right)\cr
&= \int^\infty_{0} dk f(k)  + {1\over 2} f(0)  - {1 \over 12 }  f'(0)+ {1\over 720}f'''(0) +\ldots
}}
The sum \ha\ then gives
\eqn\hc{\eqalign{
K^s_{S}(s) &= {1\over 4\pi a^2} \left[\int^\infty_{0} dk ~e^{-sk(k+1)}(2k+1)  + {1 \over 2} - {1\over 12 } (2 -s) + {1 \over 720}(-12s) + {\cal O}(s^2) \right]  \cr &= {1\over 4\pi a^2 s} \left( 1 + {1\over 3}s + {1\over 15}s^2 + \ldots\right)~.
}}
%

\subsec{The Fermion on $S^2$}
Relativistic fermions on $S^2$ transform in the $2j+1$ dimensional representations of the rotation group with half-integral values $j={1\over 2}, {3\over 2},\ldots$. The square of the Dirac operator is a scalar so it commutes with the angular momentum operator. Indeed, these operators are essentially the same (see {\it eg.} \AbrikosovNJ):
\eqn\hca{
-D_F^2=\vec{J}^2 + {1\over 4}~.
}
The eigenvalues needed for the heat kernel are thus $j(j+1) + {1\over 4} = (j+{1\over 2})^2$. Introducing the integer $k=j-{1\over 2}=0,1,\ldots$ we write the analogue of \ha\ for one fermionic degree of freedom:
\eqn\hd{\eqalign{
K^f_{S}(s) &= {1\over 4\pi a^2} \sum_{k=0}^\infty e^{-s(k+1)^2} (2k+2)={1\over 2\pi a^2} \sum_{k=0}^\infty e^{-sk^2} k~.
}}
We evaluate this expression using the Euler-MacLaurin formula \hb:
\eqn\he{\eqalign{
K^f_{S}(s) &= {1\over 4\pi a^2} \left[\int^\infty_{0} dk ~e^{-sk^2}2k  + \left(
- {1\over 12 } \cdot 2  + {1 \over 720}(-12s) + {\cal O}(s^2) \right ) \right]  \cr
& = {1\over 4\pi a^2 s} \left( 1 -  {1\over 6}s - {1\over 60}s^2 + \ldots\right) ~.
}}
We employ the convention that the heat kernel for the spinor on the sphere has the same sign as a scalar. Fermion statistics will of course ultimately change the sign of the contribution to the one loop determinant but we will take this into account manually when needed.

\subsec{Scalars and Fermions on AdS$_2$}
The expansion of the heat kernel in curvature invariants has the structure \kha\ for all fields. The only local distinction between $S^2$ and AdS$_2$ is the sign of the curvature. Further, by dimensional analysis each power of curvature is accompanied by one power of the expansion parameter $s$. Thus we can find the heat kernels on AdS$_2$ from the $S^2$ results by changing the sign of $s$. The overall sign of the heat kernel is such that the asymptotics \kg\ apply for small $s$.

Applying the $s\to -s$ rule to the scalar on $S^2$ \hc\ we find
\eqn\hf{\eqalign{
K^s_{A} (s)&= {1\over 4\pi a^2 s} \left( 1 - {1\over 3}s + {1\over 15}s^2 + \ldots\right)}~,
}
for the massless scalar on AdS$_2$. The fermion on $S^2$ \he\ similarly gives
\eqn\hg{\eqalign{
K^f_{A} (s) &=
- {1\over 4\pi a^2 s} \left( 1 + {1\over 6}s - {1\over 60}s^2 + \ldots\right)~,
}}
for each fermionic degree of freedom on AdS$_2$. We take fermion statistics into account through the overall sign in \hg.

The $s\to -s$ rule relates the local terms in the heat kernels on $S^2$ and AdS$_2$ but there are no correspondingly simple continuations of individual eigenvalues and eigenfunctions \CamporesiNW. For example, the scalar spectrum on $S^2$ is $\lambda_S = l(l+1)$ with $l=0,1,\ldots$. The scalar spectrum AdS$_2$ similarly includes a discrete branch for which $\lambda_A= - m^2 = - h(h-1)$ with $h= 1, 2,\ldots$. These highest weight type modes are important as they correspond to massive on-shell particles (in Lorentzian signature). However, the quantum fluctuations on AdS$_2$ are encoded in an unrelated continuous branch with $\lambda_A = p^2+{1\over 4}$ with $p\in R$. These are strictly off-shell modes which correspond to conformal weights $h = {1\over 2} + ip$ and ``mass'' $m^2\leq -{1\over 4}$ below the Breitenlohner-Freedman bound (for $p\neq 0$).

The expression \kc\ for a heat kernel as a ``sum" over eigenvalues in the case of AdS$_2$ becomes an integral. For a scalar field\refs{\CamporesiWN,\CamporesiGA}
\eqn\hh{
K_A^s(s) = {1\over 2\pi a^2} \int_0^\infty e^{-(p^2 + {1\over 4}) s} p \tanh\pi p ~dp = {1\over 4\pi a^2 s} \left( 1 - {1\over 3}s + {1\over 15}s^2 + \ldots\right)~.}
The Plancherel measure $\mu(p) = p \tanh\pi p $ arises as the eigenvalue space dual of the real space measure $\sqrt{-g}=\sinh\rho$ on AdS$_2$. This agrees with \hf\ as it should. The leading term for small $s$ agrees with the flat space result \kg\ both in magnitude and in sign even though this is not manifest in the prefactor of \hh\ (related to AdS$_2$ volume \kig).

\subsec{AdS$_2\times S^2$}
For minimally coupled fields the kinetic operator on the product space is a sum of kinetic operators on the factors. In this situation the eigenfunctions on the full space are products of eigenfunctions on each factor space and so the eigenvalues on the product space are equal to the sum of eigenvalues on each factor. The full Green's function \kg\ therefore becomes a product of contributions from each factor and this result descends to the heat kernel.

The heat kernel of a minimally coupled boson on AdS$_2\times S^2$ is thus
\eqn\hi{
K^s_4(s) = K^s_{S}(s)K^s_{A}(s) = {1\over 16\pi^2 a^4 s^2} \left( 1 + {1\over 45}s^2 + \ldots\right)~,
}
where the individual factors were copied from \hc\ and \hf. Similarly the heat kernel of a minimally coupled Dirac fermion on AdS$_2\times S^2$ becomes
\eqn\hj{
K^f_4(s) = 4K^f_{S}(s)K^f_{A}(s) = -{1\over 4\pi^2 a^4 s^2} \left( 1  - {11\over 180}s^2 + \ldots\right)~,
}
where the individual factors were taken from \he\ and \hg. The overall factor of $4$ counts the number of fermionic degrees of freedom. In our conventions the overall minus sign came from AdS$_2$ \hg\ but not from the $S^2$ \he. This correctly accounts for statistics on AdS$_2\times S^2$.

An important benchmark in the following section will be the heat kernel of a full hypermultiplet with {\it no} couplings taken into account. This is the heat kernel of four scalars and one Dirac fermion (with four fermionic degrees of freedom), all minimally coupled:
\eqn\hk{
K^{\rm min}_4(s)  = 4K^s_4(s) + K^f_4(s) = {1\over 4\pi^2 a^4 s^2} \cdot {1\over 12} s^2~.
}
In this case the divergences cancel to two leading orders, both of order $s^{-2}$ and of order $s^{-1}$. Thus quantum corrections do not induce a cosmological constant, nor a renormalization of the Newton constant. The leading nontrivial term in the heat kernel is constant, corresponding to a marginal operator in the action. This order is responsible for the logarithmic corrections to black hole entropy that we are interested in.

\newsec{Quantum Corrections to ${\cal N}=2$ multiplets}
The supergravity fields propagating in the AdS$_2\times S^2$ background interact with each other, in addition to the interaction with the background. This modifies their heat kernels from the canonical values such as those given in \hi\ and \hj. In this section we combine the quantum numbers computed in section 2 with the elementary methods from section 3 to determine the quantum corrections with interactions taken into account.

\subsec{The Hypermultiplet}
The classical spectrum in \bd\ gives the eigenvalues of scalars in the hypermultiplet as four towers with $(h,j)=(k+1,k)$ with $k=0,1,\ldots$. From the AdS$_2$ point of view these are on-shell particles with mass level $m^2=h(h-1)=k(k+1)$ and degeneracy $2k+1$ from an $SU(2)$ quantum number.

The AdS$_2$ heat kernels presented in \hf\ and \hg\ are for massless particles ($h=1$) with unit degeneracy but AdS$_2$ mass and degeneracy due to $SU(2)$ spin $j$ present a minimal modification
\eqn\qa{
K_A(h,j;s) = K_A (h=1,j=0;s) ~e^{-h(h-1)s} (2j+1)~.
}
The heat kernel for the four towers with $(h,j)=(k+1,k)$ therefore becomes
\eqn\qb{\eqalign{
K^{H,b}_4(s)  & =   4\cdot K^s_A (s)\cdot {1\over 4\pi a^2} \sum_{k=0}^\infty  e^{-sk(k+1)} (2k+1)  \cr
& = 4K^s_A (s)\cdot K^s_S (s)\cr
& ={1\over 4\pi^2 a^4 s^2} \left( 1 + {1\over 45}s^2 + \ldots\right)~.
}}
The sum over particles in AdS$_2$ reduced to \ha\ which was evaluated already in \hc\ where it was interpreted as the heat kernel in $S^2$.

Although in this section we take an AdS$_2$ perspective, the final result \qb\ agrees with \hi\ for four massless scalars in AdS$_2\times S^2$. This is expected because the scalar fields in hypermultiplets interact only minimally with the background. The absence of scalar couplings in turn is well known from the fact that the attractor
mechanism in the AdS$_2\times S^2$ background applies to scalars in vector multiplets but not to those in hypermultiplets \FerraraIH.

The fermions in a hypermultiplet are more complicated because couplings to the graviphoton background introduces effective masses. For a fermion the dictionary between conformal weight and spacetime mass is $m^2=h(h-1) + {1\over 4} = (h-{1\over 2})^2$ with the shift of ${1\over 4}$ the $SL(2)$ analogue of the $SU(2)$ shift in \hca. The AdS$_2$ heat kernel for the two towers of hypermultiplet fermions in \bd\ then gives
\eqn\qc{\eqalign{
K^{H,f}_4(s) & =   K^f_A(s) \cdot {1\over 4\pi a^2} \sum_{k=0}^\infty \left( e^{-sk^2} (2k+2)  + e^{-s(k+1)^2} 2k\right)\cr
& = K^f_A(s)\cdot {1\over 2\pi a^2}  \left( \sum_{k=0}^\infty  e^{-s(k+1)^2} (2k+2) + 1 \right)\cr
&= K_A^f(s)\cdot {1\over 2\pi a^2 s} \left( 1  - {1\over 6}s - {1\over 60}s^2 + \ldots + s \right)\cr
&= - {1\over 4\pi^2 a^4 s^2} \left( 1  - {11\over 180}s^2 + \ldots + s(1 + {1\over 6}s)+ \ldots  \right)~.
}}
The second line was obtained by a simple shift of indices and the third line used the summation formula \hd-\he. In the final line we used the AdS$_2$ heat kernel \hg. We refrained from collecting all terms in the final result in order to stress that the first set of terms are the ``kinematical" (not due to interactions) contributions present even for non-interacting fermions (as in \hj) while the second set of terms can be attributed to the interactions between the fermions.

The heat kernel for the full hypermultiplet is obtained by the addition of contributions from bosons \qa\ and fermions \qc:
\eqn\qd{
K^{H}_4(s) =
{1\over 4\pi^2 a^4 s^2} \left( {1\over 12}s^2 - (s + {1\over 6}s^2) + \ldots\right)=
{1\over 4\pi^2 a^4} \left( -{1\over s} - {1\over 12} + \ldots\right)~.
}
In the first form we recognize the first term as the canonical (non-interacting) result \hk\ and so the second one can be attributed to the interactions. In the context of logarithmic corrections to the area law we focus on the {\it constant} term in \qd. It is amusing that the role of the interactions for this term is precisely to {\it change the sign} of the quantum corrections. Such an effect could conceivably go unnoticed in some circumstances. Our result agrees (of course) with that reported by A. Sen \SenBA.

\subsec{The Vectormultiplet}
For the ${\cal N}=2$ vector multiplet it is well-known that the bosonic degrees of freedom are sensitive to the interactions: the attractor mechanism determines the horizon values of the scalar fields in terms of the charges of the vector fields. Thus the excitations of the scalar fields in vector multiplets acquire a mass in AdS$_2$. This should be contrasted with the scalar fields in hypermultiplets which remain freely specifiable in the near horizon region as they are moduli.

The effect of interactions on the heat kernel for the bosons in the vector multiplet are captured again by the
spectrum \bd\ which we take into account using \qa. This gives
\eqn\qe{\eqalign{
K^{V,b}_4(s) & =   2\cdot K^s_A(s)\cdot {1\over 4\pi a^2} \sum_{k=0}^\infty \left( e^{-sk(k+1)} (2k+3)  + e^{-s(k+1)(k+2)} (2k+1)\right)\cr
& = 2\cdot K^s_A(s)\cdot {1\over 2\pi a^2}  \left( \sum_{k=0}^\infty  e^{-sk(k+1)} (2k+1) + {1\over 2} \right)\cr
&= 2\cdot K^s_A(s)\cdot
{1\over 2\pi a^2 s} \left( 1  + {1\over 3}s +  {1\over 15}s^2 + \ldots + {1\over 2}s \right)\cr
&= {1\over 4\pi^2 a^4 s^2} \left( 1  + {1\over 45}s^2 + \ldots + {1\over 2}s(1- {1\over 3}s)+\ldots \right)~.
}}
The second line was obtained by a simple shift of summation indices and the third line used the evaluation of \ha\ given in \hc. The heat kernel for a scalar in AdS$_2$ was given in \hf.

According to \bd\ the four {\it fermionic} degrees of freedom are, in contrast to the bosons, minimally coupled. The contribution of the fermions to the heat kernel is therefore captured by the AdS$_2\times S^2$ result \hj\
\eqn\qf{
K^{V,f}_4(s)  = - {1\over 4\pi^2 a^4 s^2} \left( 1 - {11\over 180}s^2 + \ldots\right)~.
}

Adding \qe\ and \qf\ we find the result for the ${\cal N}=2$ vector multiplet
\eqn\qg{
K^V_4(s) = {1\over 4\pi^2 a^4 s^2} \left( {1\over 12} s^2 + {1\over 2}s(1- {1\over 3}s)\right)
= {1\over 4\pi^2 a^4} \left( {1\over 2s}- {1\over 12}\right)~.
}
Again the ``${1\over 12}s^2$" is the benchmark contribution that one gets from four fermions and four bosons
in the AdS$_2\times S^2$ background before interactions are taken into account. The ``${1\over 2}s(1- {1\over 3}s)$" can thus be attributed to the couplings between the bosons in the vector multiplet, the same interactions that give rise to the attractor mechanism for ${\cal N}=2$ black holes. The effect of interactions on the constant term in the heat kernel is to flip its sign.

\subsec{The Gravitino Multiplet}
Combining the spectrum of the fermions in \bd\ with the rule \qa\ we find the heat kernel
\eqn\ho{\eqalign{
K^{3/2,f}_4(s)  &= 2\cdot K^f_A(s)\cdot {1\over 4\pi a^2} \sum_{k=0}^\infty \left( e^{-s(k+1)^2} (2k+4)  + e^{-s(k+2)^2} (2k+2)\right)\cr
& = 2\cdot K^f_A(s)\cdot  {1\over 2\pi a^2}   \sum_{k=0}^\infty  e^{-s(k+1)^2} (2k+2) \cr
&= 2\cdot K_A^f(s)\cdot  {1\over 2\pi a^2 s} \left( 1 - {1\over 6}s - {1\over 60}s^2 + \ldots\right)\cr
&= - {1\over 4\pi^2 a^4 s^2} \left( 1 - {11\over 180}s^2 + \ldots\right)~.
}}
The summation is the same as for the minimal fermion \he. There are contributions from interactions in intermediate steps but they ultimately cancel each other.

The quantum numbers of the bosons in \bd\ are shifted relative to free bosons. The effect of this shift is to remove the leading term in the sum over modes on the sphere, which is easily taken into account:
\eqn\hp{\eqalign{
K^{3/2,b}_4(s)  & = K^s_A(s)\cdot {1\over 4\pi a^2} \sum_{k=1}^\infty  e^{-sk(k+1)} (2k+1)\cr
&= K^s_A(s)\cdot {1\over 4\pi a^2} \left(\sum_{k=0}^\infty  e^{-sk(k+1)} (2k+1) -1 \right)\cr
&=  K^s_A(s)\cdot {1\over 4\pi a^2 s} \left( 1 + {1\over 3}s + {1\over 15}s^2 + \ldots - s+\ldots \right)\cr
& = {1\over 4\pi^2 a^4 s^2} \left( 1 + {1\over 45}s^2 + \ldots - s(1 - {1\over 3})s+\ldots \right)~.
}}
The ``$-s(1- {1\over 3}s)$" can be attributed to the couplings between components of a vector field relative to those of scalar degrees of freedom.

Adding \ho\ and \hp\ we find the heat kernel for a complete ${\cal N} = 2$ multiplet for a massive gravitino:
\eqn\hpa{
K^{3/2}_4  = {1\over 4\pi^2 a^4 s^2} \left( {1\over 12}s^2   - s(1 - {1\over 3}s)\right)= {1\over 4\pi^2 a^4} \left( - {1\over s} + {5\over 12}\right)~.
}
%

\subsec{The Graviton Multiplet}
The quantum numbers $(h,j)=(k+{5\over 2},k+{3\over 2})$ from \bd\ give the contribution from the four fermion degrees of freedom as
\eqn\ho{\eqalign{
K^{{\rm grav},f}_4(s)  &= 4\cdot K^f_A(s)\cdot {1\over 4\pi a^2} \sum_{k=0}^\infty  e^{-s(k+2)^2} (2k+4) \cr
&= 4\cdot K^f_A(s)\cdot {1\over 4\pi a^2} \left( \sum_{k=0}^\infty e^{-s(k+1)^2} (2k+2) - 2e^{-s} \right) \cr
&= 4\cdot K^f_A(s)\cdot {1\over 4\pi a^2s} \left( 1 - {1\over 6}s - {1\over 60} s^2 - 2se^{-s} \right) \cr
&= - {1\over 4\pi^2 a^4 s^2} \left( 1 - {11\over 180}s^2 + \ldots - 2s ( 1  - {5\over 6}s) + \ldots \right)~.
}}
As in previous cases the ``$-2s( 1 -  {5\over 6}s)$" can be attributed to the couplings between components of a gravitino field relative to those of a free fermion.

Finally, inserting the quantum numbers \bd\ for bosons in the supergravity multiplet into \qa\ we find
\eqn\hp{\eqalign{
K^{{\rm grav},b}_4(s)  & = K^s_A(s)\cdot {1\over 4\pi a^2} \sum_{k=0}^\infty  \left( e^{-s(k+2)(k+1)} (2k+5) + e^{-s(k+3)(k+2)} (2k+3)\right) \cr
&= K^s_A(s)\cdot {1\over 4\pi a^2} \left( 2\sum_{k=0}^\infty   e^{-s(k+2)(k+1)} (2k+3) - e^{-2s}\right)\cr
&= K^s_A(s)\cdot {1\over 2\pi a^2} \left( \sum_{k=0}^\infty   e^{-s(k+1)} (2k+1) - 1 - {1\over 2}e^{-2s}\right)\cr
&=  K^s_A(s)\cdot {1\over 2\pi a^2 s} \left( 1 + {1\over 3}s + {1\over 15}s^2 + \ldots - {3\over 2}s+s^2+\ldots \right)\cr
& = {1\over 4\pi^2 a^4 s^2} \left( 1 + {1\over 45}s^2 +\ldots -{3\over 2}s + {3\over 2}s^2+\ldots \right)
}}

Adding \ho\ and \hp\ the complete result for the heat kernel of the ${\cal N}=2$ gravity multiplet becomes
\eqn\hpa{\eqalign{
K^{\rm grav}_4(s)  &= {1\over 4\pi^2 a^4 s^2} \left( {1\over 12}s^2   -{3\over 2}s(1 - {1\over 3}s) + 2s ( 1  + {1\over 6}s) \right) \cr
& = {1\over 4\pi^2 a^4 s^2} \left( {1\over 12}s^2   + ( {1\over 2}s -  {1\over 6}s^2) \right)
= {1\over 4\pi^2 a^4} \left( {1\over 2s}  -  {1\over 12}\right)~.
}}
%

\subsec{Summary}
In summary, we have computed the contributions to heat kernels of the ${\cal N}\geq 2$ theory from physical non-zero modes. The result is
\eqn\hq{
K_{\rm nzm} ={1\over 4\pi^2 a^4} \left(  ( {1\over 2s} -{1\over 12}) + ({\cal N}-2)(-{1\over s} + {5\over 12}) + n_V ({1\over 2s} - {1\over 12})+ n_H ( - {1\over s} - {1\over 12})\right)
}
The notation ``nzm" is a reminder that at this point interactions have been taken into account but the focus was on non-zero modes. Corrections due to zero-modes will be considered in the next two sections.

\newsec{Boundary States}
As we have stressed, the spectrum \bd\ enumerates physical modes only. In particular, gauge conditions have been imposed that fix the gauge symmetry. These conditions remove all unphysical states except that, for each continuous gauge symmetry, a single physical boundary mode remains. We discussed the mechanism for this in some detail in section 2.

The physical boundary states contribute to the quantum corrections to black holes just like all other physical states.
In this section we compute their contributions to the heat kernel.

\subsec{Localization on the Boundary}
A 4D gauge symmetry reduces to a tower of 2D gauge symmetries in AdS$_2$. Each entry in the tower gives rise to a single mode on the boundary of AdS$_2$. These towers were presented as a list in \db.

The contribution from each entire tower will amount to a field on the $S^2$ that is localized on AdS$_2$. We need to find the spectrum of these fields on $S^2$. This can be accomplished by considering the structure of gauge transformations. This introduces gauge dependence at intermediate stages but our final result is gauge invariant.

In the following we consider the boundary modes for each ${\cal  N}=2$ multiplet in turn.

\medskip

\noindent
{\it The Vector Multiplet}

Modes that are pure gauge from the 4D point of view take the form of a gauge variation
\eqn\hmaa{
\delta {\cal A}_I = \nabla_I \Lambda~,
}
where $\Lambda$ is the $U(1)$ gauge parameter. Among these modes those that preserve the Lorentz gauge
condition
\eqn\hma{
\nabla_I {\cal A}^I=0~,
}
are
\eqn\hmb{
-\nabla^I \delta{\cal A}_I = -\nabla^2 \Lambda =0 ~,
}
just like a massless scalar from the 4D point of view. From the 2D point of view there is a tower of fields in AdS$_2$ with masses given by
\eqn\hmbb{
m^2=k(k+1)~,
}
with $k=0,1,\ldots$. Each field is pure gauge so its contribution to physical processes cancels with the corresponding unphysical mode. This cancellation is imperfect and leaves the AdS$_2$ zero-mode $\nabla_A^2\Lambda=0$. We interpret this mode as a physical mode on the AdS$_2$ boundary. As we recombine all 2D fields $k=0,1,\ldots$ we find a physical scalar field on $S^2$.
The quantum corrections due to these physical states are computed by the scalar determinant on the sphere \hc\  and gives
\eqn\hmc{
K^V_{\rm bndy}=  {1\over 2\pi a^2}\cdot {1\over 4\pi a^2 s} \left(1 + {1\over 3} s \right)
= {1\over 4\pi^2 a^4} \left( {1\over 2s}  + {1\over 6} \right)~.
}
The overall factor is the volume of AdS$_2$. The sign is the one appropriate for a physical boson.
The simple pole in the parameter $s$ is mild for a 4D field but entirely standard for a 2D field.

\medskip

\noindent
{\it The Gravitino Multiplet}

The gauge symmetry of a gravitino is the SUSY variation
\eqn\hmca{
\delta \Psi_I = \nabla_I \epsilon ~.
}
The SUSY transformation that preserves the Lorentz gauge condition on the gravitino
\eqn\hmd{
\gamma^I \delta \Psi_I=0~,
}
satisfies the Weyl's equation
\eqn\hme{
\gamma^I \nabla_I \epsilon =0~.
}
The physical boundary state that remains is therefore a Weyl fermion on $S^2$. Our previous computation of the heat kernel for a
single fermionic degree of freedom \he\ then gives
\eqn\hmf{
K^{(3/2)}_{\rm bndy} = - {1\over 2\pi a^2} \cdot {1\over 4\pi a^2 s} \cdot 2\left(1 - {1\over 6} s \right)
= {1\over 4\pi^2 a^4} \left( - {1\over s}  + {1\over 6} \right)~.
}
An explicit factor of two counted the two components of the Weyl fermion. The overall minus sign is appropriate for a physical fermion.

The gravitino supermultiplet also includes two vector multiplets. Each realizes a standard $U(1)$ gauge symmetry and gives rise to
a boundary mode that contributes \hmc\ to the heat kernel. The total boundary contribution to the gravitino supermultiplet therefore becomes
\eqn\hmp{
K^{(3/2)}_{\rm bndy} = {1\over 4\pi^2 a^4} \cdot {1\over 2} ~.
}
There is no pole in $s$ because the boundary states in this multiplet fill out a super multiplet with equal number of fermions and bosons on the boundary.

\medskip

\noindent
{\it The Graviton Multiplet}

The gauge symmetries of gravity are the 4D diffeomorphisms $\xi^I$ acting on gravitational perturbations as
\eqn\hrb{
\delta h_{IJ} = \nabla_I \xi_J + \nabla_J \xi_I~.
}
The coordinate transformations that preserve the Lorentz (harmonic) gauge condition
\eqn\hra{
\nabla^I h_{\{IJ\}} = \nabla^I (h_{IJ}+ h_{JI} - g_{IJ} h_K^K) = 0~,
}
satisfy
\eqn\hrb{
 (g_{IJ}\nabla^2 + R_{IJ}) \xi^J =0~.
}
The Ricci curvature is $R_{\mu\nu} = -g_{\mu\nu}$ on AdS$_2$ and $R_{\alpha\beta} = +g_{\alpha\beta}$ on the $S^2$.

The diffeomorphisms $\xi^\alpha$ generate vector modes on $S^2$ so the angular momentum of the corresponding boundary modes is restricted to $k=1,2,\ldots$. The Ricci curvature gives a contribution $\Delta m^2=-1$ to the effective mass and the dualization to a scalar field gives an identical contribution. The spectrum of the two scalar boundary modes with $\nabla^2_A\xi^\alpha=0$ therefore becomes
\eqn\hrbb{
m^2_S = k(k+1) - 2 ~,
}
with $k=1,\ldots$. The mass-shift $\Delta m^2=-2$ is such that the leading AdS$_2$ boundary mode is massless also on the $S^2$.

The pure gauge modes generated by $\xi^\mu$ decompose into an AdS$_2$ scalar $\nabla_\mu \xi^\mu$, an AdS$_2$ vector
$\nabla_\mu\xi_\nu - \nabla_\nu\xi_\mu$, and an AdS$_2$ traceless tensor. The AdS$_2$ scalar mixes with the pure gauge mode from the graviphoton such that all three of these are independent even though $\xi^\mu$ has only two components. The
AdS$_2$ zero-modes of the scalar and the traceless tensor both give rise to physical boundary states with the spectrum \hmbb\ of a standard scalar field on $S^2$. However, the AdS$_2$ vector has zero modes that generates a tower of boundary modes with the shifted effective mass
\eqn\hrba{
m^2_S=k(k+1)+2~.
}
These three towers all have $k=0,\ldots$. The leading terms with vanishing angular momentum $j=k=0$ are essentially the boundary states denoted $\delta X$ and $\delta \rho$ in section 2.2 except that here the gauge is different and the graviphoton is taken into account.

The sum of contributions from all five bosonic boundary modes yields
\eqn\hmu{\eqalign{
K_{\rm bndy}^{\rm grav, b} & =  2\cdot {1\over 4\pi^2 a^4}\cdot {1\over 2}  \left( {1\over s} - {2\over 3}\right)e^{2s} +
2\cdot {1\over 4\pi^2 a^4} \cdot {1\over 2}\left( {1\over s} + {1\over 3}\right) +   {1\over 4\pi^2 a^4}\cdot {1\over 2} \left( {1\over s} + {1\over 3}\right)e^{-2s}
\cr
& = {1\over 4\pi^2 a^4} \cdot {5\over 2} ({1\over s} + {1\over 3})~.
}}
Despite the various shifts of masses and angular momentum quantum numbers this is identical to the heat kernel of five free scalars on the $S^2$.


The ${\cal N}=2$ supersymmetry acts on the two gravitini in the graviton multiplet as
\eqn\hha{
\delta \Psi^A_I = (\delta^A_B \nabla_I -  {1\over 4}{\hat F}\epsilon_{AB}\gamma_I )\epsilon^B~,
}
where the background graviphoton fieldstrength ${\hat F} = {1\over 2} F_{JK}\gamma^{JK} = \epsilon_{\alpha\beta} \gamma^{\alpha\beta}$.
This differs from a generic gravitino \hmca\ by the dependence on the graviphoton background. It is because of this dependence that ${\cal N}=2$ SUSY is preserved. The field strength contributions to \hha\ are such that the AdS$_2$ ground state energy $(-\nabla_\mu \nabla^\mu)$ of the two fermions adds to $\Delta m^2=-1$. This gives a shift in the effective fermion mass on $S^2$ such that
\eqn\hhb{
m^2 = (k+1)^2 - 1~.
}
The first term is the standard effective mass \hca\ on $S^2$, sometimes written as $j(j+1)+{1\over 4} = (j+{1\over 2})^2$ with $j$ taking half integer values. The tower $j={1\over 2}, {3\over 2},\ldots$ is parametrized here by $k=0,1,\ldots$. The mass-shift $\Delta m^2=-1$ is such that the leading AdS$_2$ boundary mode is massless also on the $S^2$.

The heat kernel for a single standard fermion on $S^2$ was given in \he. Four fermionic boundary degrees of freedom with effective
mass \hhb\ then give
\eqn\cf{
K^{\rm grav, f}_{bndy}  =  - 4\cdot {1\over 4\pi^2 a^4}\cdot {1\over 2} \left(  {1\over s} - {1\over 6} \right)e^{s} = - {1\over 4\pi^2 a^4}\left(  {2\over s} + {5\over 3} \right)~.
}
Adding the bosonic contribution \hmu\ we have
\eqn\hmr{
K^{\rm grav}_{\rm bndy} = {1\over 4\pi^2 a^4} \left({1\over 2s}  - {5\over 6}\right)~,
}
for the complete contribution of boundary states to the heat kernel of the ${\cal N}=2$ supergravity multiplet.

\medskip

\noindent
{\it Summary}

In summary, the contribution to the heat kernel of the ${\cal N}=2$ theory from boundary modes is
\eqn\hqv{
K_{\rm bndy} ={1\over 4\pi^2 a^4} \left(  ( {1\over 2s}  - {5\over 6} ) + ({\cal N}-2)\cdot {1\over 2} + n_V ({1\over 2s} + {1\over 6}) \right)~.
}

We can add this to the bulk contribution \hq\ and find
\eqn\hqw{
K_{\rm phys} ={1\over 4\pi^2 a^4} \left(  ( {1\over s}  - {11\over 12} ) + ({\cal N}-2)\cdot ( -{1\over s}  + {11\over 12} )  +
n_V ({1\over s} + {1\over 12})  + n_H ( - {1\over s}  - {1\over 12}) \right)~.
}
As a nontrivial consistency check on \hqw\ note that the coefficient of $1/s$ is the same for each type of ${\cal N}=2$ multiplet, except that the sign alternates as the spin of the SUSY multiplet changes. This is precisely the property needed to ensure that these terms cancel in any theory with ${\cal N}=4$ SUSY, as they should.

Another interesting special case is the pure ${\cal N}=3$ theory which is scale invariant at this level \NicolaiTD. The ${\cal N}=3$ matter multiplets have $n_H=n_V=1$ so an arbitrary number of those can be added without violating scale invariance.

\newsec{Zero-Modes}
The boundary states are zero modes from the AdS$_2$ point of view but they are generally non-trivial on the $S^2$. The true 4D zero-modes are the boundary states that are also zero modes on the $S^2$. These zero mode contributions require special considerations.

The zero mode content of each multiplet can be read off from the spectrum of boundary states. The vector multiplet has one bosonic zero-mode from gauge symmetry: the $k=0$ entry in \hmbb. The gravitino multiplet has two bosonic zero-modes, both from gauge symmetry. The gravity multiplet also has two bosonic zero-modes: the $k=1$ entry in \hrbb. These both have angular momentum $j=1$. Finally, the gravity multiplet also has four fermionic zero-modes, the $k=0$ entry in \hhb.

For the zero-modes we cannot use the Euclidean path integral \ka\  (repeated here for easy reference)
\eqn\kaw{
e^{-W} = \int {\cal D}\phi ~e^{-\phi\Lambda\phi} = {1\over\sqrt{{\rm det}\Lambda}}~,
}
since they correspond to vanishing eigenvalues of the matrix $\Lambda$. However, each zero-mode is just a field in zero dimensions so in this sector the path integral reduces to an ordinary integral. The scale dependence of $N_0$ zero-modes with scaling dimension $\Delta$ is
\eqn\kax{
e^{-W} = \int {\cal D}\phi_0 = {\rm Vol} [\phi_0] \sim \epsilon^{-N_0\Delta}~.
}
In contexts where \kaw\  applies it is understood that the dependence on physical parameters is encoded in {\it ratios} of integrals of this general form. The scale dependence due to a single zero-mode is similarly computed from ratios of integrals \kax\ computed at different scales.

The na\"{\i}ve inclusion of $N_{\rm 0}$ zero-modes in the heat kernel \kc:
\eqn\ia{
D(s) = \sum_i e^{-s\lambda_i} = \sum_{\lambda_i\neq0}e^{-s\lambda_i} + N_0~,
}
corresponds to a term $W = N_0 \ln \epsilon$ in the effective action according to \kb. Thus the correct zero-mode contribution $W\sim \Delta N_0 \ln\epsilon$ from \kax\ is larger than the na\"{\i}ve result by a factor of the scaling dimension $\Delta$. After generalization to multiple fields with either bosonic or fermionic statistics we have
\eqn\ic{
K_{\rm zm} =  {1\over 8\pi^2 a^4}\sum_{i\in B} N_{0,i}(\Delta_i - 1) - {1\over 8\pi^2 a^4}\sum_{i\in F} N_{0,i}(2\Delta_i - 1)~,
}
for the correction to the heat kernel $K(s)$ due to zero-modes. Each fermionic zero-mode counts with double weight because of the leading spin degeneracy in \hd.

Vector fields have dimension $\Delta_1=1$ so they were already taken correctly into account in the na\"{\i}ve heat kernel. Since
the zero-modes in the vector and (massive) gravitino multiplet are all due to vector fields these multiplets do not get corrected.
It is only the supergravity-multiplet that is corrected due to zero-modes.

Disregarding the vector, the bosonic zero-modes in the gravity multiplet are just $k=1$ in \hrbb. Each of these two states have angular momentum $j=1$ so there are $N_0^b=2\cdot (2j+1)=6$ bosonic zero-modes in the path integral. These fields have scaling dimension $\Delta_2=2$. Similarly, \hhb\ gives $N_0^f=4$ fermionic zero-modes in the path integral. They have scaling dimension $\Delta_{3/2}={3\over 2}$. The zero-mode contribution to a general ${\cal N}=2$ theory simply becomes
\eqn\if{
K_{zm} = {1\over  8\pi^2 a^4}  \cdot ( 6\cdot (2-1) - 4\cdot (3- 1) )
= {1\over  4\pi^2 a^4}  \cdot ( 3- 4 ) = {1\over  4\pi^2 a^4} \left( - 1\right)~.
}

\subsec{Summary}
The sum of contributions to the heat kernel from non-zero modes \hq, boundary modes \hqv, and zero modes \if\ is
\eqn\iu{
K_{\rm tot} ={1\over 4\pi^2 a^4} \left(  ( {1\over s}  - {23\over 12}) + ({\cal N}-2)(-{1\over s} + {11\over 12}) + n_V ({1\over s} + {1\over 12})+ n_H ( - {1\over s} - {1\over 12})\right)~.
}
This is the main result of our computations.

\newsec{Logarithmic Corrections to the Black Hole Entropy}
In this section we give a brief but self-contained review of the relation between the heat kernel and the quantum corrections to the black hole entropy.

\subsec{The Trace Anomaly}
The trace of the energy momentum tensor including quantum corrections can be divided into a divergent term and a finite (renormalized) term
\eqn\khg{
T^\mu_{\mu, {\rm tot}} = T^\mu_{\mu, {\rm div}} + T^\mu_{\mu, {\rm ren}}~.
}
Each of these terms is related to an analogous term in the effective action as
\eqn\kiz{
T^\mu_{\mu}  = {2\over\sqrt{-g}} g^{\mu\nu} {\delta W\over\delta g^{\mu\nu}}~.
}
In even dimensions the heat kernel takes the form
\eqn\kia{
D(s)= {\rm sing.} + D_0 + {\cal O}(s)~,
}
where ``sing.'' indicates terms with poles at $s=0$ while $D_0$ is the constant that encodes the trace anomaly. According to \kb\ the constant $D_0$ corresponds to the logarithmically divergent term
\eqn\kib{
W_{\rm div}\sim {1\over 2}D_0\ln\epsilon^2~,
}
in the effective action.

In theories with classical scale invariance $T^\mu_{\mu, {\rm tot}} =0$ and so
\eqn\kj{
T^\mu_{\mu, {\rm ren}} = - T^\mu_{\mu, {\rm div}} = - {2\over\sqrt{-g}} g^{\mu\nu} {\delta W_{\rm div}\over\delta g^{\mu\nu}} = {2\over {\rm Vol}}{\partial W_{\rm div}\over\partial\ln\epsilon^2}
= {1\over {\rm Vol}}D_0~.
}
We can still use this result for the anomaly in theories without classical scale invariance. Of course such theories have, in addition, a classical (non-anomalous) contribution to the trace of the energy momentum tensor. The volume factor is again the regulated volume exhibited in \kig.

\subsec{The Black Hole Entropy}
For extremal black holes the entropy $S= - W_{\rm ren}$ and so the logarithmic dependence of the entropy
is determined by
\eqn\kk{
{\partial S\over\partial\ln A_H} = - {\partial W_{\rm ren}\over\partial\ln a^2}
= - {1\over 2} \int d^D x T^\mu_{\mu, {\rm ren}}  = - {1\over 2}D_0~.
}
The dependence on the physical scale $\ln a$ and the UV cut-off scale $\epsilon$ has the opposite sign. The result for the logarithmic correction to the entropy therefore becomes
\eqn\kl{
\delta S = {1\over 2}D_0\ln A_H = 4\pi^2 a^4 K_0\ln A_H~,
}
where $K_0$ is the constant term in the heat kernel density \ki.

The relation \kk\ between the trace anomaly and the logarithmic correction to the entropy is interesting and quite general. It is corrected only by the treatment of zero-modes. Our formula \ic\ for the zero-mode contribution to the heat kernel was constructed precisely so that the entropy formula \kl\ would be maintained for this contribution as well.

The constant term in the heat kernel expansion $K_0$ is easily read off from the total heat kernel \iu. The relation \kl\ then gives the logarithmic correction so the entropy
\eqn\km{\eqalign{
\delta S = {1\over 12} \left(23 - 11({\cal N}-2)- n_V + n_H  \right)\ln A_H~.
}}
This is the final result advertised in the introduction as \aa.

\medskip

\medskip

\noindent
{\bf Acknowledgement}

We thank A. Sen for discussions. This work was supported in by the US Department of Energy under grant DE-FG02-95ER40899.

\listrefs

\end